\documentclass[1p,times]{elsarticle}

\usepackage{ecrc}

\usepackage{url}

\volume{00}

\firstpage{1}

\journalname{Physics Procedia}

\runauth{}

\jid{phpro}

\jnltitlelogo{Physics Procedia}

\usepackage{amssymb}

\usepackage[figuresright]{rotating}
\usepackage{graphicx}

\begin{document}

\begin{frontmatter}



\dochead{}

\title{Diagnosis and Prediction of Tipping Points in Financial Markets: Crashes and Rebounds}


\author[ETH]{Wanfeng Yan},
\ead{wyan@ethz.ch}
\author[ETH]{Ryan Woodard},
\ead{rwoodard@ethz.ch}
\author[ETH,SFI]{Didier Sornette\corref{cor}}
\cortext[cor]{Corresponding author. Address: KPL F 38.2, Kreuzplatz
5, ETH Zurich, CH-8032 Zurich, Switzerland. Phone: +41 44 632 89 17,
Fax: +41 44 632 19 14.}
\ead{dsornette@ethz.ch}%
\ead[url]{http://www.er.ethz.ch/}%

\address[ETH]{Chair of Entrepreneurial Risks, Department of
  Management, Technology and Economics, ETH Zurich, CH-8001 Zurich,
  Switzerland}%
\address[SFI]{Swiss Finance Institute, c/o University of Geneva, 40
  blvd. Du Pont dArve, CH 1211 Geneva 4, Switzerland}%

\begin{abstract}
By combining (i) the economic theory of rational expectation bubbles,
(ii) behavioral finance on imitation and herding of investors and
traders and (iii) the mathematical and statistical physics of
bifurcations and phase transitions, the log-periodic power law (LPPL)
model has been developed as a flexible tool to detect bubbles. The LPPL
model considers the faster-than-exponential (power law with finite-time
singularity) increase in asset prices decorated by accelerating
oscillations as the main diagnostic of bubbles. It embodies a positive
feedback loop of higher return anticipations competing with negative
feedback spirals of crash expectations. 
The power of the LPPL model is illustrated by two recent real-life
predictions performed recently by our group: the peak of the Oil
price bubble in early July 2008 and the burst of a bubble on the Shanghai
stock market in early August 2009. We then present the concept
of ``negative bubbles'', which are the mirror images of positive bubbles.
We argue that similar positive feedbacks are at work to fuel
these accelerated downward price spirals. We adapt
the LPPL model to these negative bubbles and implement 
a pattern recognition method to predict the end times of 
the negative bubbles, which are characterized by rebounds
(the mirror images of crashes associated with the standard
positive bubbles). The out-of-sample tests quantified by
error diagrams demonstrate the high significance of the
prediction performance.
\end{abstract}


\end{frontmatter}


\section{Introduction}
\label{Introduction}

Bubbles and crashes in financial markets are of global significance because of
their effects on the lives and livelihoods of a majority of the world's
population.  In spite of this, the science to correctly identify bubbles in
advance of their associated crashes produces fewer successful results than that
used to treat baldness, choose `quality' videos to watch or find a date on the
internet.  Instead, pundits and experts alike line up after the fact to claim
that a particular bubble was obvious in hindsight.  We present here
a report on recent progress of the on-going effort of our group, the 
Financial Crisis Observatory at ETH Zurich (\url{www.er.ethz.ch/fco/},
to advance the science of understanding why, how and when bubbles form so that
they can be identified \textit{before} they spectacularly crash and spread
misfortune to those who knowingly or unknowingly had bet on a long position.

The basis for our approach contradicts the accepted wisdom of the Efficient
Market Hypothesis, which claims that large deviations from fundamental prices
(i.e., bubbles and crashes) only exist when a new piece of information drops
(exogenously) onto an unsuspecting market on a very short time scale.  Instead,
we claim that bubbles are the result of endogenous market dynamics over a much
longer time scale--weeks, months and years.  Because of the long build-up of
these effects, bubbles can be identified by particular dynamical signatures
predicted by our theoretical framework.

Among the well-documented history of financial bubbles and crashes over the
past 400 years, through countless significantly different countries and kings,
empires and economies, regulations and reform, there has been one consistent
ingredient in all booms and busts: humans.  Only human behavior has survived
all attempts at preventing repeats in the wake of disastrous crashes.  Much of
the dynamics of the long time scales mentioned above is due to humans acting
like humans: those without the knowledge imitate one another in the absence of
a clearly better alternative and take refuge in the comfort of the crowd
(herding) while those with the knowledge refute the masses and claim these
noise traders are wrong.

We hypothesize that the signatures of this characteristic human behavior can be quantitatively
identified.  The imitation and herding behavior creates positive feedback in
the valuation of an asset, resulting in a greater-than-exponential (power law)
growth of the price time series.  The tension and competition between the
learned experts and the noise traders creates decorations on this power law
growth comprising oscillations that are periodic in the logarithm of time.
Log-periodic oscillations appear to our clocks as peaks and valleys with
progressively smaller amplitudes and greater frequencies that eventually reach
a point of no return, where the unsustainable growth has the highest
probability of ending in a violent crash or gentle deflation of the bubble.

The log periodic power law (LPPL) model has thus been developed as a 
flexible tool to detect bubbles. This model combines (i) the economic
theory of rational expectation bubbles,
(ii) behavioral finance on imitation and herding of investors and
traders and (iii) the mathematical and statistical physics of
bifurcations and phase transitions.
The LPPL model considers the faster-than-exponential (power law with finite-time
singularity) increase in asset prices decorated by accelerating
oscillations as the main diagnostic of bubbles. It embodies a positive
feedback loop of higher return anticipations competing with negative
feedback spirals of crash expectations. 

The power of the LPPL model is illustrated by several recent real-life
predictions performed by our group. We will present two examples in this paper: one is the peak of the Oil
price bubble in early July 2008 and the other is the burst of a bubble on the Shanghai
stock market in early August 2009. We then present the concept
of ``negative bubbles'', which are the mirror images of positive bubbles.
We argue that similar positive feedbacks are at work to fuel
these accelerated downward price spirals. We adapt
the LPPL model to these negative bubbles and implement 
a pattern recognition method to predict the end times of 
the negative bubbles, which are characterized by rebounds
(the mirror images of crashes associated with the standard
positive bubbles). The out-of-sample tests quantified by
error diagrams demonstrate the high significance of the
prediction performance.

This paper is constructed as follows.
We will give a general introduction to the LPPL model in section \ref{lppl} . In section \ref{case}, two successful real-life predictions mentioned above are discussed. Then we describe
the phenomenon of negative bubbles and their termination in the form of rebounds. We develop
a pattern recognition method based on the LPPL model and test its prediction performance in section  \ref{pr}. 
Section \ref{conclusion} concludes.

\section{Log-periodic Power Law}
\label{lppl}

\subsection{Long time scale fermentation of bubbles}
\label{longscale}
In sharp contrast to the Efficient Market Hypothesis (EMH) that crashes result from novel negative
information incorporated in prices at short time scales, we build on the
radically different hypothesis summarized by Sornette \cite{Sornettecrash},
that the underlying causes of the crash should be found in the preceding
year(s).  We define a bubble as a market regime in which
the price accelerates ``super-exponentially''.  The term ``super-exponential'' 
means that the growth rate of the price grows itself.  A constant
price growth rate (also called return) leads to an exponential growth,
the normal average trajectory of most economic and financial time series.
When the growth rate grows itself, the price accelerate hyperbolically.
This growth of the growth rate is interpreted as being due to 
progressively increasing build-up of market
cooperation between investors.  As the bubble matures, 
its approaches a critical point, at which an instability can be triggered
in the form of a crash, or more generally a change of regime.
This critical point can also be called a phase transition, a bifurcation,
a catastrophe or a tipping point. According to this
``critical'' point of view, the specific manner by which prices collapse is not
the most important problem: a crash occurs because the market has entered an
unstable phase and any small disturbance or process may reveal the existence of
the instability. Think of a ruler held up vertically on your finger: this very
unstable position will lead eventually to its collapse as a result of a small
(or an absence of adequate) motion of your hand or due to any tiny whiff of
air. The collapse is fundamentally due to the unstable position; the
instantaneous cause of the collapse is secondary. In the same vein, the growth
of the sensitivity and the growing instability of the market close to such a
critical point might explain why attempts to unravel the proximal origin of the
crash have been so diverse. Essentially, anything would work once the system is
ripe.

What is the origin of the maturing instability?  A follow-up hypothesis
underlying this proposal is that, in some regimes, there are significant
behavioral effects underlying price formation leading to the concept of
``bubble risks''. This idea is probably best exemplified in the context of
financial bubble, where, fuelled by initially well-founded economic
fundamentals, investors develop a self-fulfilling enthusiasm by an imitative
process or crowd behavior that leads to the building of castles in the air, to
paraphrase Malkiel \cite{Malkiel90}.

Our previous research suggests that the ideal economic view, that stock markets
are both efficient and unpredictable, may be not fully correct. We propose
that, to understand stock markets, one needs to consider the impact of positive
feedbacks via possible technical as well as behavioral mechanisms such as
imitation and herding, leading to self-organized cooperation and the
development of possible endogenous instabilities. We thus propose to explore
the consequences of the concept that most of the crashes have fundamentally an
endogenous, or internal, origin and that exogenous, or external, shocks only
serve as triggering factors. As a consequence, the origin of crashes is
probably much more subtle than often thought, as it is constructed
progressively by the market as a whole, as a self-organizing process. In this
sense, the true cause of a crash could be termed a systemic instability.

\subsection{Imitation and Herding Among Humans as the Cause of Bubbles}
\label{imiherd}

Humans are perhaps the most social mammals and they shape their environment to
their personal and social needs. This statement is based on a growing body of
research at the frontier between new disciplines called neuro-economics,
evolutionary psychology, cognitive science and behavioral finance
(\cite{Damasio94}, \cite{Camerer03}, \cite{Gintis05}). This body of evidence
emphasizes the very human nature of humans with its biases and limitations,
opposed to the previously prevailing view of rational economic agents
optimizing their decisions based on unlimited access to information and to
computation resources.

We hypothesize that financial bubbles are footprints of perhaps the most robust
trait of humans and the most visible imprint in our social affairs: imitation
and herding (see Sornette \cite{Sornettecrash} and references
therein). Imitation has been documented in psychology and in neuro-sciences as
one of the most evolved cognitive processes, requiring a developed cortex and
sophisticated processing abilities. In short, we learn our basics and how to
adapt mostly by imitation all through our life. It seems that imitation has
evolved as an evolutionary advantageous trait, and may even have promoted the
development of our anomalously large brain (compared with other mammals),
according to the so-called ``social brain hypothesis'' advanced by R. Dunbar
\cite{dunbar}. It is actually ``rational'' to imitate when lacking sufficient
time, energy and information to make a decision based only on private
information and processing, that is, most of the time. Imitation, in obvious or
subtle forms, is a pervasive activity of humans. In the modern business,
economic and financial worlds, the tendency for humans to imitate leads in its
strongest form to herding and to crowd effects. Imitation is a prevalent form
in marketing with the development of fashion and brands.
Cooperative herding and imitation lead to positive feedbacks, that is, an
action leads to consequences which themselves reinforce the action and so on,
leading to virtuous or vicious circles. 

The methodology that we have developed
consists in using a series of mathematical and computational
formulations of these ideas, which capture the hypotheses that (1) bubbles can
be the result of positive feedbacks and (2) the dynamical signature of bubbles
derives from the interplay between fundamental value investment and more
technical analysis. The former can be embodied in nonlinear extensions of the
standard financial Black-Scholes model of log-price variations
\cite{SorAndersen02,IdeSornette02,Corcosetal,AndersenSor04}.  The later
requires more significant extensions to account for the competition between
(i) inertia separating analysis from decisions, (ii) positive momentum feedbacks and (iii)
negative value investment feedbacks \cite{IdeSornette02}.

\subsection{Positive Feedback Among Traders Leads to Power Law Growth in Asset Price}
\label{pl}
\begin{figure}[htp]
\centering
\includegraphics[width=12cm]{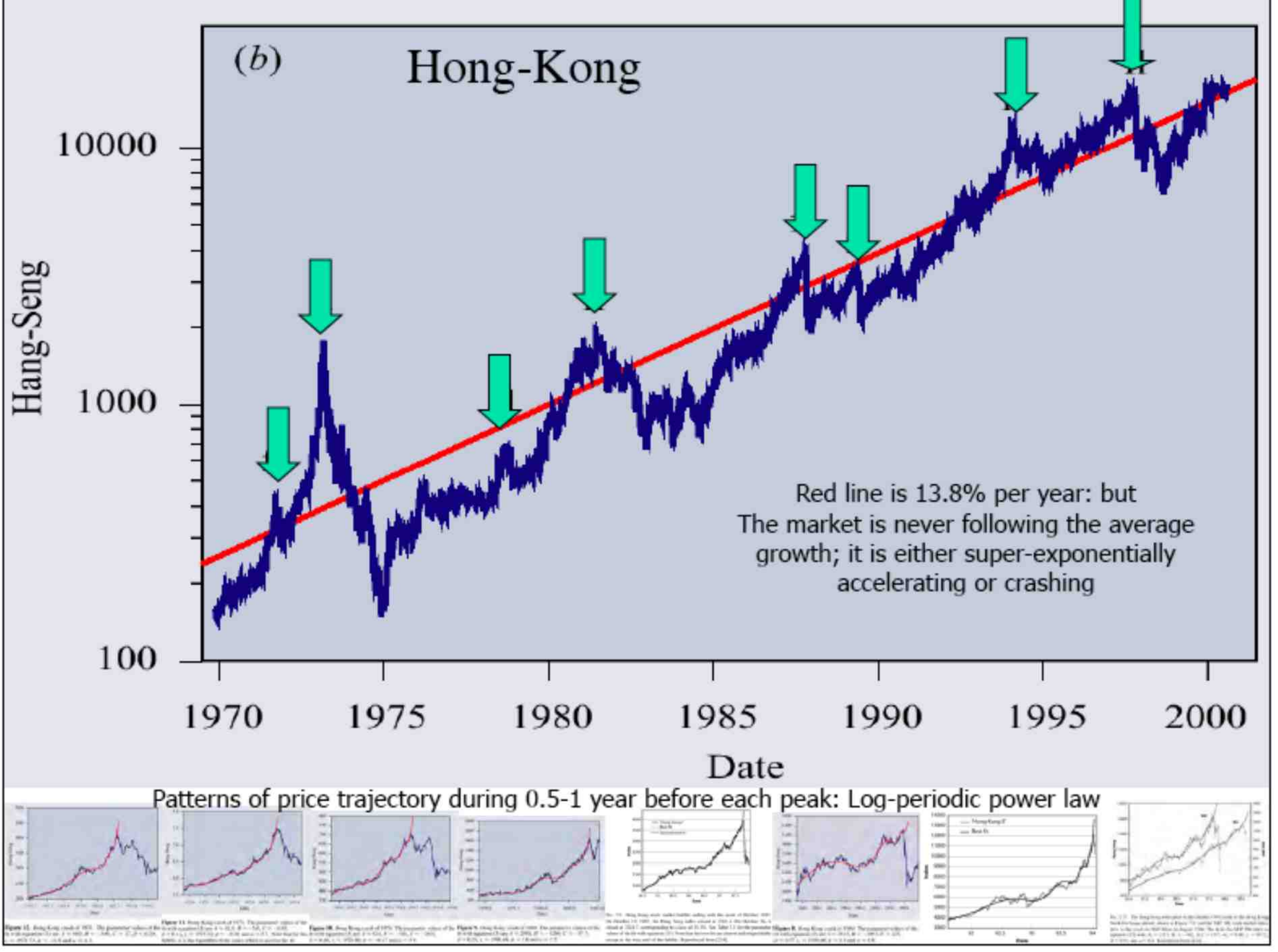}
\caption{\label{Fig:Fig_HengSeng_Bubbletextbook} Trajectory of the Hong-Kong
  Hang Seng index from 1970 to 2000. The vertical log-scale together with the
  linear time scale allows one to qualify an exponential growth with constant
  growth rate as a straight line. This is indeed the long-term behavior of this
  market, as shown by the best linear fit represented by the solid straight
  line, corresponding to an average constant growth rate of 13.8\% per
  year. The 8 arrows point to 8 local maxima that were followed by a drop of
  the index of more than 15\% in less than three weeks (a possible definition
  of a crash).  The 8 small panels at the bottom show the upward curvature of
  the log-price trajectory preceding each of these local maxima, which diagnose
  unsustainable bubble regimes, each of which culminates at its peak before
  crashing.}
\end{figure}

The idea of positive feedback has led us to propose that one of the hallmarks
of a financial bubble is the faster-than-exponential growth of the price of the
asset under consideration, as already mentioned. It is convenient to model this accelerated growth by
a power law with a so-called finite-time singularity \cite{SorTakaZhou}.  This
feature is nicely illustrated by the price trajectory of the Hong Kong Hang
Seng index from 1970 to 2000, as shown in
Fig.~\ref{Fig:Fig_HengSeng_Bubbletextbook}. The Hong Kong financial market is
repeatedly rated as providing one of the most pro-economic,
pro-entrepreneurship and free market-friendly environments in the world, and
thus provides a textbook example of the behavior of weakly regulated liquid and
striving financial markets. In Fig.~\ref{Fig:Fig_HengSeng_Bubbletextbook}, the
logarithm of the price $p(t)$ is plotted as a function of time (in linear
scale), so that an upward trending straight line qualifies as exponential
growth with a constant growth rate: the straight solid line corresponds indeed
to an approximately constant growth rate of the Hang Seng index equal to 13.8\%
per year.

The most striking feature of Fig.~\ref{Fig:Fig_HengSeng_Bubbletextbook} is not
this average behavior but instead the obvious fact that the real market is
never following and abiding to a constant growth rate. One can observe a
succession of price run-ups characterized by growth rates $\ldots$ growing
themselves: this is reflected visually in
Fig.~\ref{Fig:Fig_HengSeng_Bubbletextbook} by transient regimes characterized
by strong upward curvature of the price trajectory. Such an upward curvature in
a linear-log plot is a first visual diagnostic of a faster than exponential
growth (which of course needs to be confirmed by rigorous statistical
testing). Such a price trajectory can be approximated by a characteristic
transient finite-time singular power law of the form
\begin{equation}
  \label{eq:pl}
  \log{p(t)} = A + B (t_c - t)^m
\end{equation}
where $B<0$, $0<m<1$ and $t_c$ is the theoretical critical time corresponding
to the end of the transient run-up (end of the bubble). \textbf{Such transient
  faster-than-exponential growth of $p(t)$ is our definition of a bubble.} It
has the major advantage of avoiding the conundrum of distinguishing between
exponentially growing fundamental price and exponentially growing bubble price,
which is a problem permeating most of the previous statistical tests developed
to identify bubbles \cite{gurkaynak-2008,Lux-Sornette}. The conditions $B<0$
and $0<m<1$ ensure the super-exponential acceleration of the price, together
with the condition that the price remains finite even at $t_c$. Stronger
singularities can appear for $m<0$.

To see that faster-than-exponential growth is naturally related to positive
feedback, let us consider the following simple presentation. Consider a
population of animals of size $x$ which grows with some constant rate $k$,
i.e., $dx/dt = k \: x$. Then, growth is exponential in that $x(t) = x(0) \:
e^{kt}$. On the other hand, positive feedback in growth dynamics arises if the
growth rate $k$ itself depends on the population size $x$ in that the growth
rate $k = k(x)$ increases with the population size. A particular simple example
is the setting $k(x) \sim x^{m-1}$, where $m>1$. Indeed, $m-1$ can be regarded
as a measure of the degree of cooperation within the population: the higher $m$
is the larger is the degree of cooperation. In the case of no cooperation,
growth dynamics is exponential. Positive multiplicative feedback generates
growth which is faster than exponential. Indeed, due to growth dynamics given
by $dx/dt = k \:x^m$, the size of the population growth exhibits a finite-time
sigularity at $t_c$. This singularity is attained according to $x(t) = x(0) [1
- \tau] ^ \frac{1}{1-m}$, $0 < \tau<1$ where $\tau = \frac{t}{t_c}$. Note that
when $m = 1$, no cooperation, growth is indeed exponential and when $m>1$,
growth dynamics is in fact faster than exponential.  It is remarkable that a
critical time $t_c$ emerges apparently out of nowhere.  Actually, $t_c$ is
determined by the initial conditions and the structure of the growth
equation. This emergence of $t_c$ which depends on the initial conditions
justifies its name in mathematical textbooks as a ``movable singularity.''

Many systems exhibit similar transient super-exponential growth regimes, which
are described mathematically by power law growth with an ultimate finite-time
singular behavior.  An incomplete list of examples includes: planet formation
in solar systems by runaway accretion of planetesimals, rupture and material
failures, nucleation of earthquakes modeled with the slip-and-velocity, models
of micro-organisms interacting through chemotaxis aggregating to form fruiting
bodies, the Euler rotating disk, and so on. Such mathematical equations can
actually provide an accurate description of the transient dynamics not too
close to the mathematical singularity where new mechanisms come into play. The
singularity at $t_c$ mainly signals a change of regime. In the present context,
$t_c$ is the end of the bubble and the beginning of a new market phase,
possibly a crash or a different regime.

Such an approach may be thought at first sight to be inadequate or too naive to
capture the intrinsic stochastic nature of financial prices, whose null
hypothesis is the geometric random walk model (\cite{Malkiel90}). However, it is
possible to generalize this simple deterministic model to incorporate nonlinear
positive feedback on the stochastic Black-Scholes model, leading to the concept
of stochastic finite-time singularities 
\cite{SorAndersen02,Fogedby1,Fogedby2,AndersenSor04,LinRenSornette,LinSornette09}.
Still much work needs to be done on this theoretical aspect.

Coming back to Fig.~\ref{Fig:Fig_HengSeng_Bubbletextbook}, one can also notice
that each burst of super-exponential price growth is followed by a crash, here
defined for the eight arrowed cases as a correction of more than 15\% in less
than 3 weeks. These examples suggest that the non-sustainable super-exponential
price growths announced a ``tipping point'' followed by a price disruption,
i.e., a crash. The Hong-Kong Hang Seng index shows that the average exponential
growth of the index is punctuated by a succession of bubbles and crashes, which
seem to be the norm rather than the exception.

\subsection{Competition between different types of traders lead to log-periodic oscillations}
\label{lp}
More sophisticated models than Eq.~(\ref{eq:pl}) have been proposed to take
into account the interplay between technical trading and herding (positive
feedback) versus fundamental valuation investments (negative mean-reverting
feedback). Accounting for the presence of inertia between information gathering
and analysis on the one hand and investment implementation on the other hand
\cite{Farmer02}, and taking additionally into account the
coexistence of trend followers and value investing
\cite{IdeSornette02}, the resulting price dynamics develop second-order
oscillatory terms and boom-bust cycles. Value investing does not necessarily
cause prices to track value. Trend following may cause short-term trend in
prices but, together with value investing and inertia, also causes longer-term oscillations.

The simplest model generalizing (\ref{eq:pl}) and including these ingredients
is the so-called log-periodic power law (LPPL) model (see Sornette
\cite{Sornettecrash} and references therein). Formally, some of the
corresponding formulas can be obtained by considering that the exponent $m$ is
a complex number with an imaginary part, where the imaginary part expresses the
existence of a preferred scaling ratio $\gamma$ describing how the continuous
scale invariance of the power law (\ref{eq:pl}) is partially broken into a
discrete scale invariance \cite{DSI-sornette98}.  The LPPL structure may also
reflect the discrete hierarchical organization of networks of traders, from the
individual to trading floors, to branches, to banks, to currency blocks. More
generally, it may reveal the ubiquitous hierarchical organization of social
networks recently reported \cite{Zhou-Dunbar05} to be associated with the
social brain hypothesis \cite{dunbar}.

Examples of calibrations of financial bubbles with one implementation of the
LPPL model are the 8 super-exponential regimes discussed above in
Fig.~\ref{Fig:Fig_HengSeng_Bubbletextbook}: the 8 small insets at the bottom of
the figure show the LPPL calibration on the Hang Seng index. Preliminary tests
\cite{Sornettecrash} suggest that the LPPL model provides a good starting point
to detect bubbles and forecast their most probable end. Rational expectation
models of bubbles a la Blanchard and Watson implementing the LPPL model
\cite{Johansen-Ledoit1,Johansen-Ledoit2} have shown that the end of the bubble
is not necessarily accompanied by a crash, but it is indeed the time where a
crash is the most probable. But crashes can occur before (with smaller
probability) or not at all. That is, a bubble can land smoothly, approximately
one-third of the time, according to preliminary investigations
\cite{Johansen-Sornette}.  Therefore, only probabilistic forecasts can be
developed. Probability forecasts are indeed valuable and commonly used in daily
life, such as in weather forecast.

\section{Successful case studies}
\label{case}

\subsection{The Oil Bubble of 2008}
\label{oil}

In \cite{sorwoodzhou}, we have presented the prediction performed
ex-ante and its post-mortem analysis of the bubble
that has developed on oil prices in USD and in other major
  currencies. The ex-ante diagnostic of the oil bubble
  was performed on the basis of the identification of unsustainable faster-than-exponential
  behavior during the course of 2007 until the mid-2008.  
  We found support for the hypothesis that the oil price
  run-up in the first half of 2008 was amplified by speculative behavior of the type found
  during a bubble-like expansion. We also attempted to unravel the
  information hidden in the oil supply-demand data reported by two
  leading agencies, the US Energy Information Administration (EIA) and
  the International Energy Agency (IEA). We suggested that the found
  increasing discrepancy between the EIA and IEA figures provides a
  measure of the estimation errors. Rather than a clear transition to
  a supply restricted regime, we interpreted the discrepancy between the
  IEA and EIA as a signature of uncertainty. This is compatible
  with the idea that there is no better
  fuel than uncertainty to promote speculation. Our post-crash
  analysis confirmed that the oil peak in July 2008 occurred within the
  expected 80\% confidence interval predicted ex-ante with data available in
  our pre-crash analysis.
  
  \begin{figure}
\centering
  \includegraphics[width=14cm]{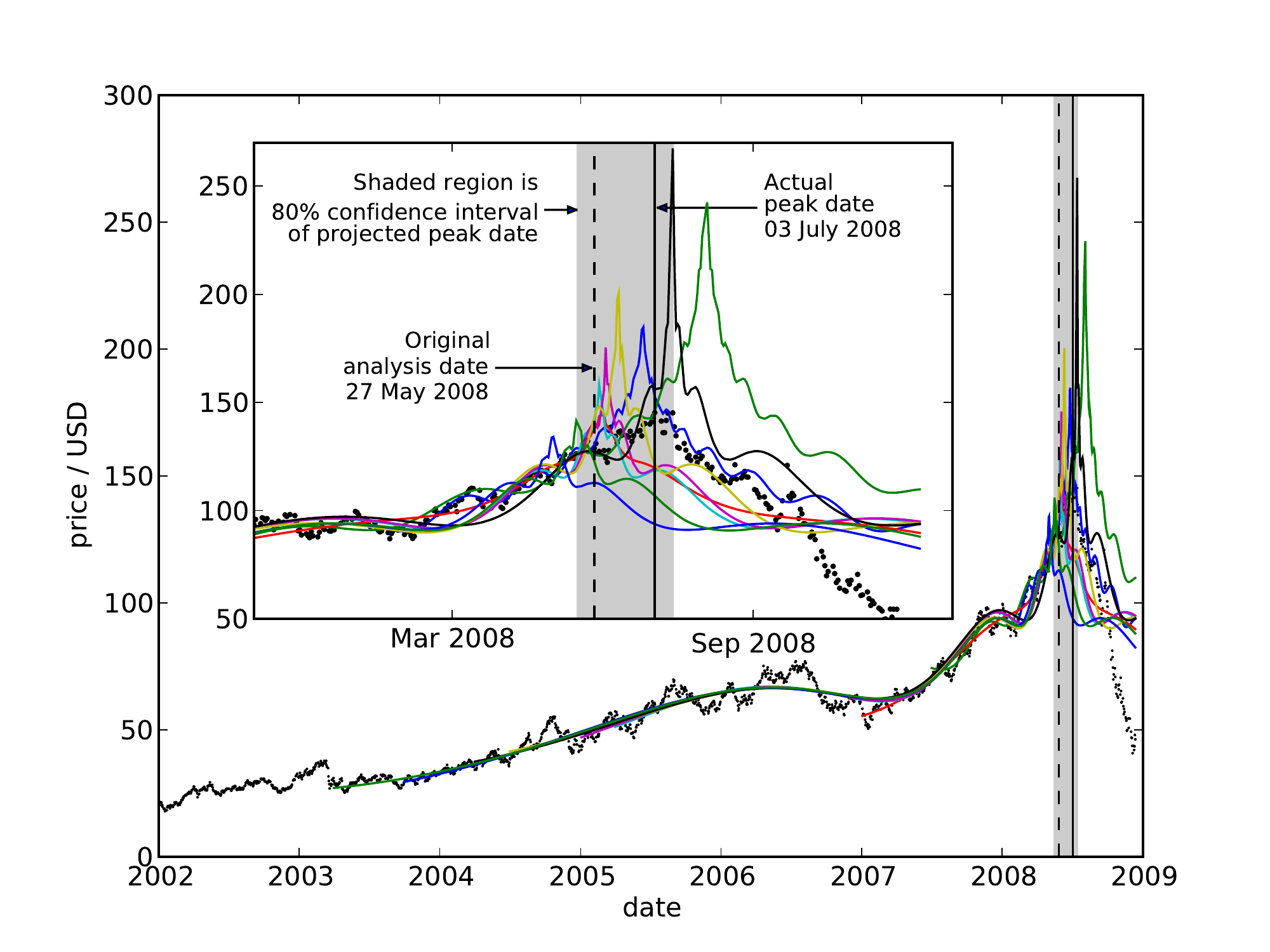}
  \caption{Time series of observed prices in USD of ``NYMEX Light
    Sweet Crude, Contract 1'' from the Energy Information
    Administration of the U.S. Government (see {\protect\url{{http://www.eia.doe.gov/emeu/international/Crude2.xls}}}) and simple
    LPPL fits (see text for explanation).  The oil price time series
    was scanned in multiple windows defined by ($t_1$, $t_2$), where
    $t_1$ ranged from 1~April~2003 to 2~January~2008 in steps of
    approximately 3 months (see text) and $t_2$ was fixed to
    23~May~2008.  Shaded box shows the 80\% confidence interval of fit
    parameter $t_c$ for fits with $t_c$ less than six months beyond
    $t_2$.  Also shown are dates of our original analysis in June~2008
    and the actual observed peak oil price on 3~July~2008. Reproduced from 
    \protect\cite{sorwoodzhou}.}
  \label{fig:oil_orig_fits}
\end{figure}
  
 The main result of our post-crash analysis is shown in
Fig.~\ref{fig:oil_orig_fits}.  We
calculated the 80\% confidence interval for the critical time $t_c$ of
the end of the Oil bubble, which was found to be
17~May~2008 to 14~July~2008 and is shown as the shaded box in the main
and inset plots of Fig.~\ref{fig:oil_orig_fits}.  The actual peak oil
price was observed to be 3~July and the steep descent in price began
on 11~July.  Both dates are within the confidence interval calculated
with the LPPL model using data through the last week of May.  This
confirms that the simple LPPL model was a successful predictor of the
2008 oil price bubble.

\subsection{The Chinese Index Bubble of 2009}
\label{shanghai}

In the midst of the global financial crisis of the past year, when the prices
of many assets and indexes fell, the Shanghai Composite Index defied financial
gravity and continued climbing as it had done since October 2008.

On 10 July 2009 (using data through 9 July 2009), we submitted our prediction
online to \url{arXiv.org} \cite{bastiaensen-china-2009}, in which we gave the
20\%/80\% (respectively 10\%/90\%) quantiles of the projected crash dates to be
17-27 July 2009 (respectively 10 July - 10 August 2009). This corresponds to a
60\% (respectively 80\%) probability that the end of the bubble occurs and that
the change of regime starts in the interval within those time windows.  Redoing
the analysis 5 days later with data through 14 July 2009, the predictions
tightened up with a 80\% probability for the change of regime to start between
19 July and 3 August 2009 (unpublished).  Our forecasts were featured in the
press at, among other places,
\url{http://www.technologyreview.com/blog/arxiv/23839/}.

On 29 July 2009, Chinese stocks suffered their steepest drop since November
2008, with an intraday bottom of more than 8\% and an open-to-close loss of
more than 5\%. The market rebounded with a peak on 4 August 2009 before
plummeting the following weeks. The Shanghai Index slumped 22 percent in
August, the biggest decline among 89 benchmark indices tracked world wide by
Bloomberg, in stark contrast with being the best performing index during the
first half of 2009. We thus successfully predicted time windows for
this crash in advance with the same
methods used to successfully predict the peak in mid-2006 of the US
housing bubble \citep{Zhou-Sornette-2006b-PA} and the peak in July 2008
of the global oil bubble \citep{sorwoodzhou}.  The
more recent bubble in the Chinese indexes was detected and its end or
change of regime was predicted independently by two groups with similar
results, showing that the model has been well-documented and can be
replicated by industrial practitioners. 

In \cite{jiangetal},
we presented a thorough post-crash analysis of this 2008-2009 Chinese
bubble and, also, the previous 2005-2007 Chinese bubble in
Ref.~\cite{jiang-china-2009}.  This publication also documents another original
forecast of the 2005-2007 bubble (though there was not a publication on that,
except a public announcement at a hedge-fund conference in Stockholm in October
2007).  Also, it clearly lays out some of our many technical methods used in testing
our analyses and forecasts of bubbles: the search and fitting method of the
LPPL model itself, Lomb periodograms of residuals to further identify the
log-periodic oscillation frequencies, (H, q)-derivatives
\cite{Zhou-Sor02Hq,Zhou-Sor03Hq} and, most recently, unit root tests of the
residuals to confirm the Ornstein-Uhlenbeck property of their stationarity
\cite{LinSornette09}. Here, we reproduce the main
figure documenting the advance prediction which included the peak of the bubble
on 4 August 2009 in its 5-95\% confidence limits. The curves are fitted by first order Laudau model:

\begin{equation}
	\label{Eq:Landau1}
	\log{p(t)} = A + B (t_c - t)^m + C (t_c - t)^m \cos(\omega \ln (t_c - t) - \phi)
\end{equation}

\begin{figure}[htp]
\centering
\includegraphics[width=6.5cm]{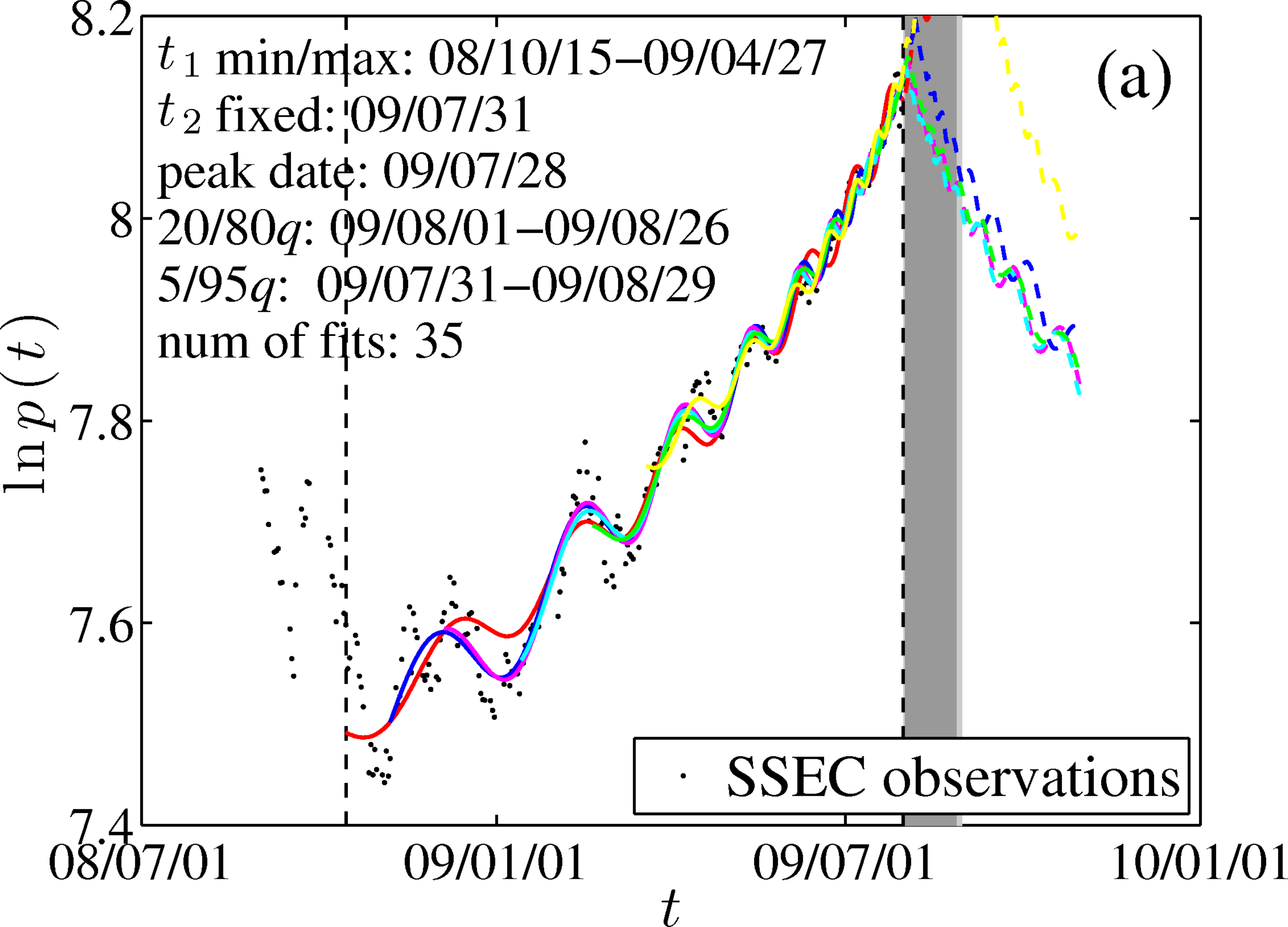}
\includegraphics[width=6.5cm]{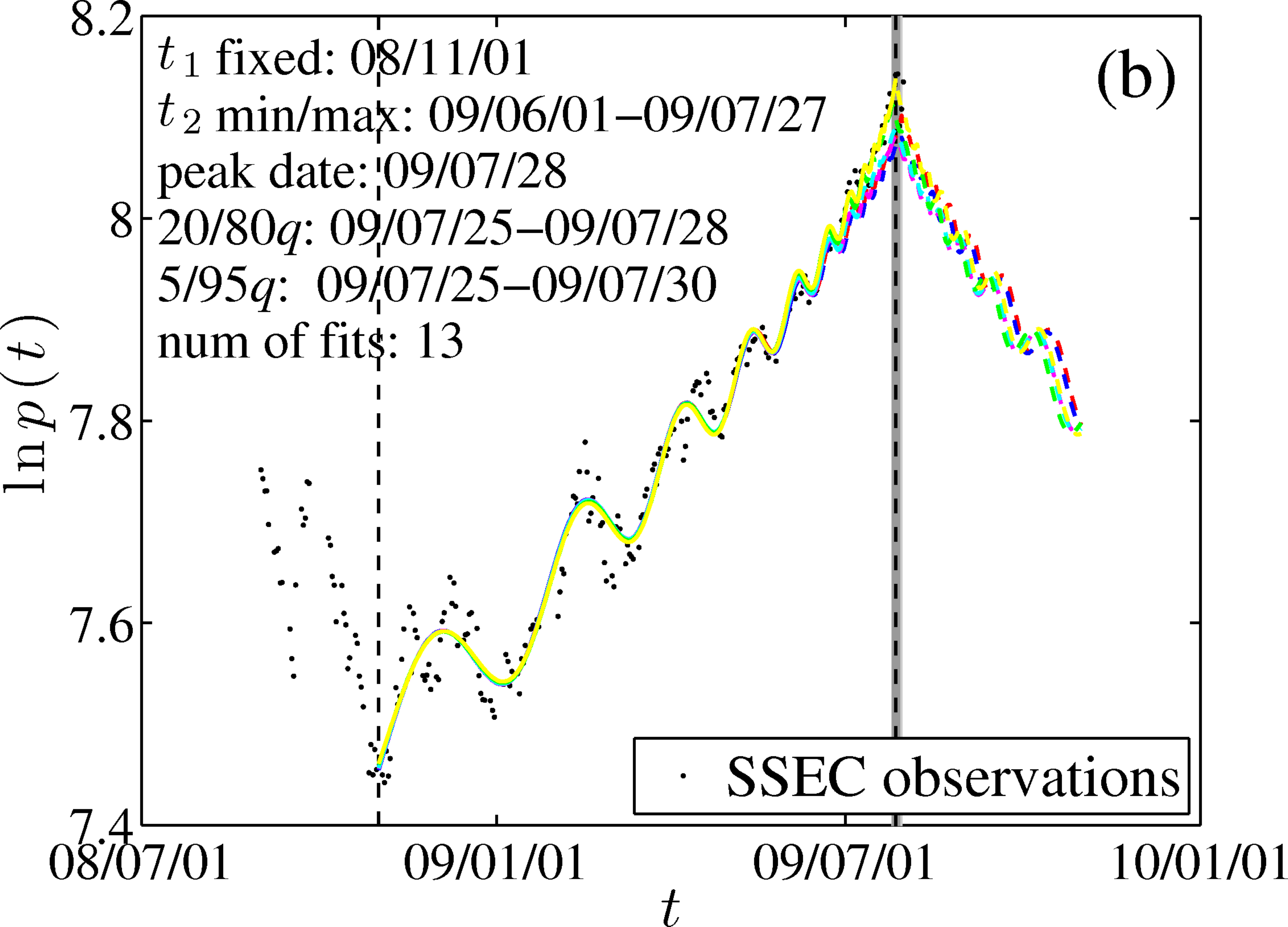}
\includegraphics[width=6.5cm]{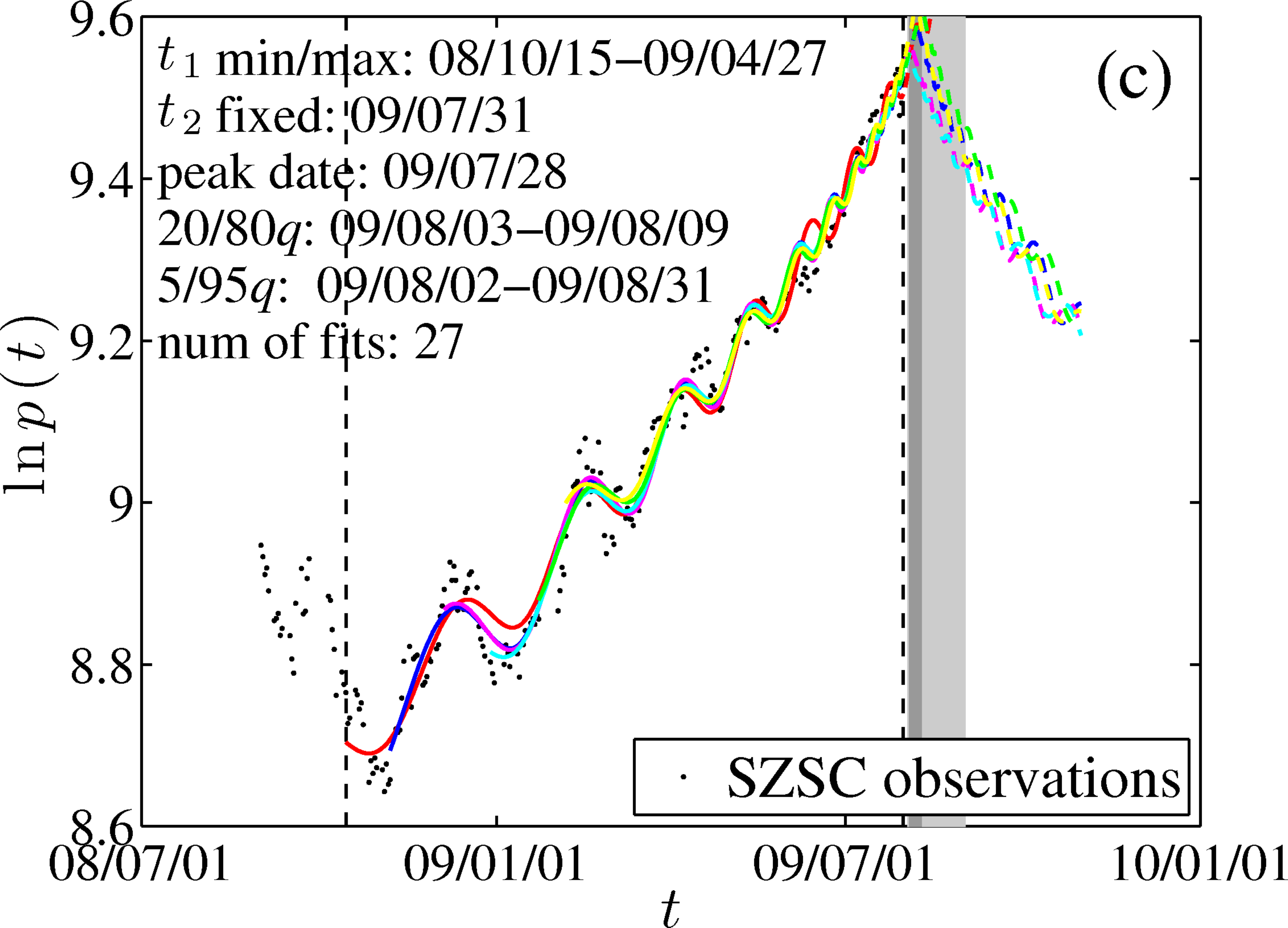}
\includegraphics[width=6.5cm]{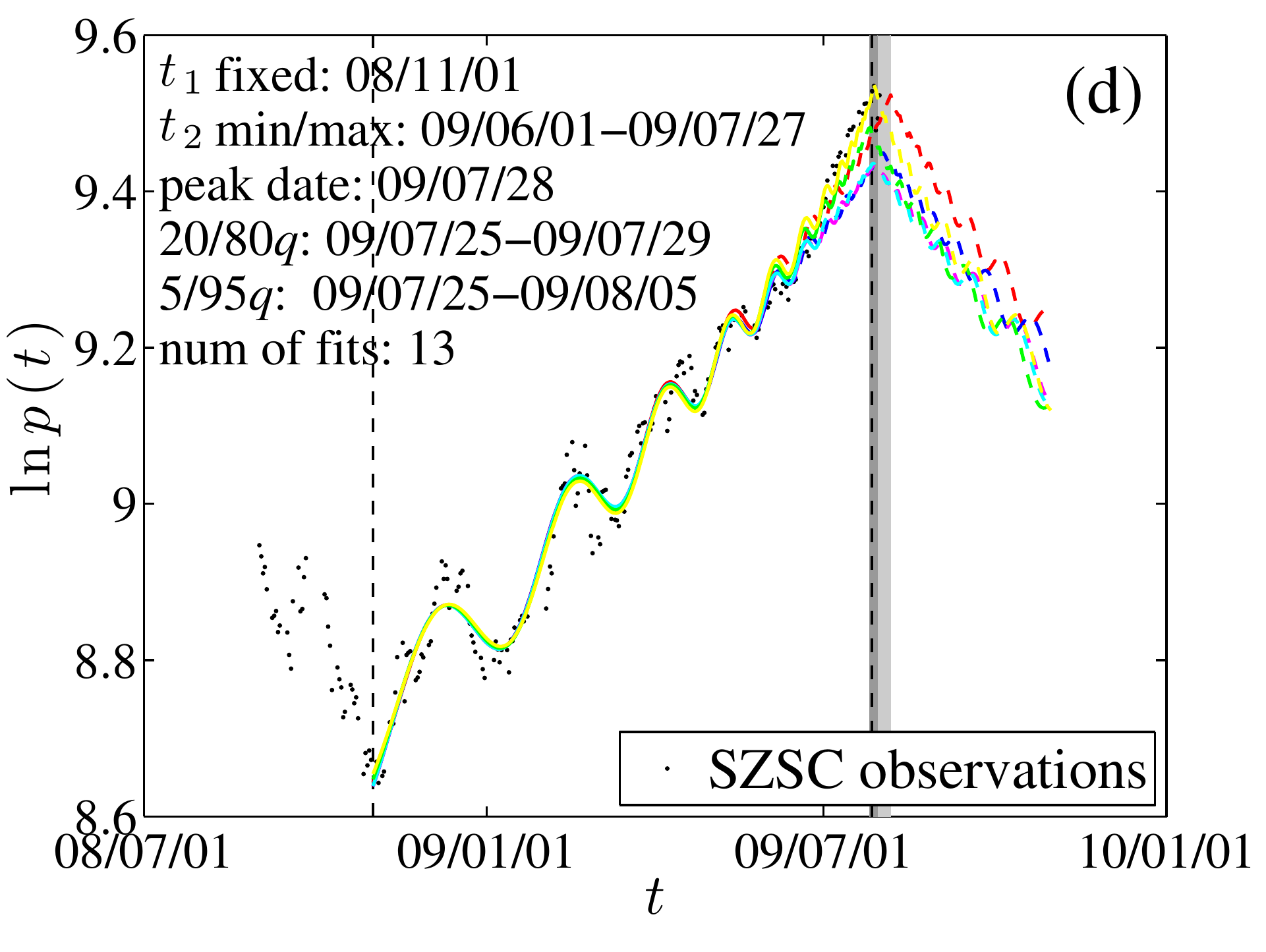}
\caption{\label{Fig:B0809} Daily trajectory of the logarithmic SSEC (a,b) and
  SZSC (c,d) index from Sep-01-2008 to Jul-31-2009 (dots) and fits to the LPPL
  formula (\ref{Eq:Landau1}). The dark and light shadow box indicate 20/80\%
  and 5/95\% quantile range of values of the crash dates for the fits,
  respectively. The two dashed lines correspond to the minimum date of $t_1$
  and the fixed date of $t_2$.  (a) Examples of fitting to shrinking windows
  with varied $t_1$ and fixed $t_2$ = Jul-31-2009 for SSEC. The six fitting
  illustrations are corresponding to $t_1$ = Oct-15-2008, Nov-07-2008,
  Dec-05-2008, Jan-05-2008, Feb-06-2008, and Feb-20-2008. (b) Examples of
  fitting to expanding windows with fixed $t_1$ = Nov-01-2008 and varied $t_2$
  for SSEC. The six fitting illustrations are associated with $t_2$ =
  Jun-01-2009, Jun-10-2009, Jun-19-2007, Jun-29-2007, Jul-13-2007,
  Jul-27-2007. (c) Examples of fitting to shrinking windows with varied $t_1$
  and fixed $t_2$ = Jul-31-2009 for SZSC. The six fitting illustrations are
  corresponding to $t_1$ = Oct-15-2008, Nov-03-2008, Nov-26-2008, Dec-19-2008,
  Jan-14-2008, and Jan-23-2008. (d) Examples of fitting to expanding windows
  with fixed $t_1$ = Dec-01-2005 and varied $t_2$ for SZSC. The six fitting
  illustrations are associated with $t_2$ = Jun-01-2009, Jun-10-2009,
  Jun-19-2007, Jun-29-2007, Jul-13-2007, Jul-27-2007. Reproduced from Jiang et. al. \cite{jiangetal}.}
\end{figure}

\section{Detection of Rebounds using Pattern Recognition Method}
\label{pr}

Until now, we have focused our attention on bubbles, their peaks
and the crashes that often follow. We have argued that bubbles
develop due to positive feedbacks pushing the price upward towards
an instability at which the bubble ends and the price may crash.

But positive feedbacks do not need to be active just for price run-ups, i.e.,
for exuberant ``bullish'' market phases. Positive feedbacks may also be 
at work during ``bearish'' market regimes, when pessimism
replaces optimism and investors sell in a downward spiral. Zhou and Sornette (2003)
have developed a generalized Weierstrass LPPL formulation to model
the 5 drops and their successive rebounds observed in the US
S\&P 500 index during the period from 2000 to 2002 \cite{zhousor}.

Here, we go further and develop a methodology that combines a LPPL
model of accelerating price drops, termed a ``negative bubble,'' and a 
pattern recognition method in order to diagnose the critical time $t_c$ of
the rebound (the antithesis of the crash for a ``positive bubble'').
A negative bubble is modeled with the same tools as a positive bubble,
that is, with  the power law expression (\ref{eq:pl}). But the essential
difference is that the coefficient $B$ is positive for
a negative bubble (while it is negative for a normal positive bubble, as discussed
above). The exponent
$m$ obeys the same condition $0<m<1$ as for the positive bubbles. 
The positivity of $B$ together with the condition $0<m<1$ implies
that the average log-price trajectory exhibits a downward curvature, 
expressing a faster-than exponential downward acceleration. In other
words, the log-price trajectory associated with a negative bubble
is the upside-down mirror image of the log-price trajectory of a positive bubble.
Additional log-periodic
oscillations are added to the LPPL model, which also account
for the competition between value investors and trend followers.

We adapt the pattern recognition method of \cite{gel}
to generate predictions of rebound times in financial markets.  
A similar method has been developed by Sornette and Zhou (2006)
to combine the information provided by the LPPL model with
a pattern recognition method for the prediction of the end of bubbles \cite{sorzhou}.

We analyze the
S\&P 500 index prices, obtained from Yahoo! finance for ticker `\^{ }GSPC'
(adjusted close price)\footnote{http://finance.yahoo.com/q/hp?s=\^{ }GSPC}.
The start time of our time series is 1950-01-05, which is very close to the
first day of S\&P 500 index (1950-01-03). The last day of our tested time
series is 2009-06-03.

We first divide our S\&P 500 index time series into different sub-windows
$(t_1, t_2)$ of length $dt \equiv t_2 - t_1$ according to the following rules:
\begin{enumerate}
\item The earliest start time of the windows is $t_{10} = $ 1950-01-03.  Other
  start times $t_1$ are calculated using a step size of $dt_1 = 50$ calendar
  days.
\item The latest end time of the windows is $t_{20} = $ 2009-06-03.  Other end
  times $t_2$ are calculated with a negative step size $dt_2 = -50$ calendar
  days.
\item The minimum window size is $dt_{\mathrm{min}} = 110$ calendar days.
\item The maximum window size is $dt_{\mathrm{max}} = 1500$ calendar days.
\end{enumerate}
These rules lead to 11,662 windows in the S\&P 500 time series.

Then we fit the log-price time series
in each window with the LPPL model and get the corresponding parameters. 

A rebound is defined by:
\begin{equation}
  \mathrm{Rbd} = \{d \mid P_d = \min\{P_x\}, \forall x \in [d-200,d+200]\}
  \label{eq:rbd}
\end{equation}
where $P_d$ is the adjusted closing price on day $d$.  According to this definition,
a rebound is said to have occurred on day $d$ if condition (\ref{eq:rbd}) holds.

The prediction method uses a learning in-sample data set. Then, an out-of-sample dataset allows
us to test the validity and success of the prediction method.
We take all the fits before 1975-01-01 as learning set. In the learning set, if the critical time of a fit is near some of the rebounds, then we select this fit into Class I. In contrast, if there is not any rebound near the critical time of a fit, this fit will be classified into Class II. We then construct the statistics of all the parameters of the fits from Class I and Class II separately. We then compare the properties of these two sets of statistics for the two classes. 
The statistically significant differences between these properties allow us to 
define the features which are specific to each class. 

Using these features associated with each class, we analyze every trading day before 1975-01-01. 
We collect all the fits with a critical time near this trading day, and find out the features of these fits. If most of the features in these fits belong to Class I, we interpret this as evidence that this trading day has a high probability to be a rebound. Otherwise, this trading day is not likely to be a rebound. 

To quantify the probability that this trading day is a rebound, we develop the rebound alarm index as follows:
\begin{equation}
  RI_t = \left \{ \begin{array}{ll}
      \frac{\nu_{t,I}}{\nu_{t,I}+\nu_{t,II}}& \mbox{if
      $\nu_{t,I}+\nu_{t,II} \geq 0$}\\
      0& \mbox{if $\nu_{t,I}+\nu_{t,II} = 0$}
                  \end{array}
                  \right.
\end{equation}
where $\nu_{t, I}, \nu_{t, II}$ represent the numbers of features of Class I and Class II respectively. 
From this definition, we can see that $RI_t \in [0,1]$. If $RI_t$ is high,
then we can declare that this day has a high probability that the rebound will start.

\begin{figure}[htp]
\centering
\includegraphics[width=\textwidth]{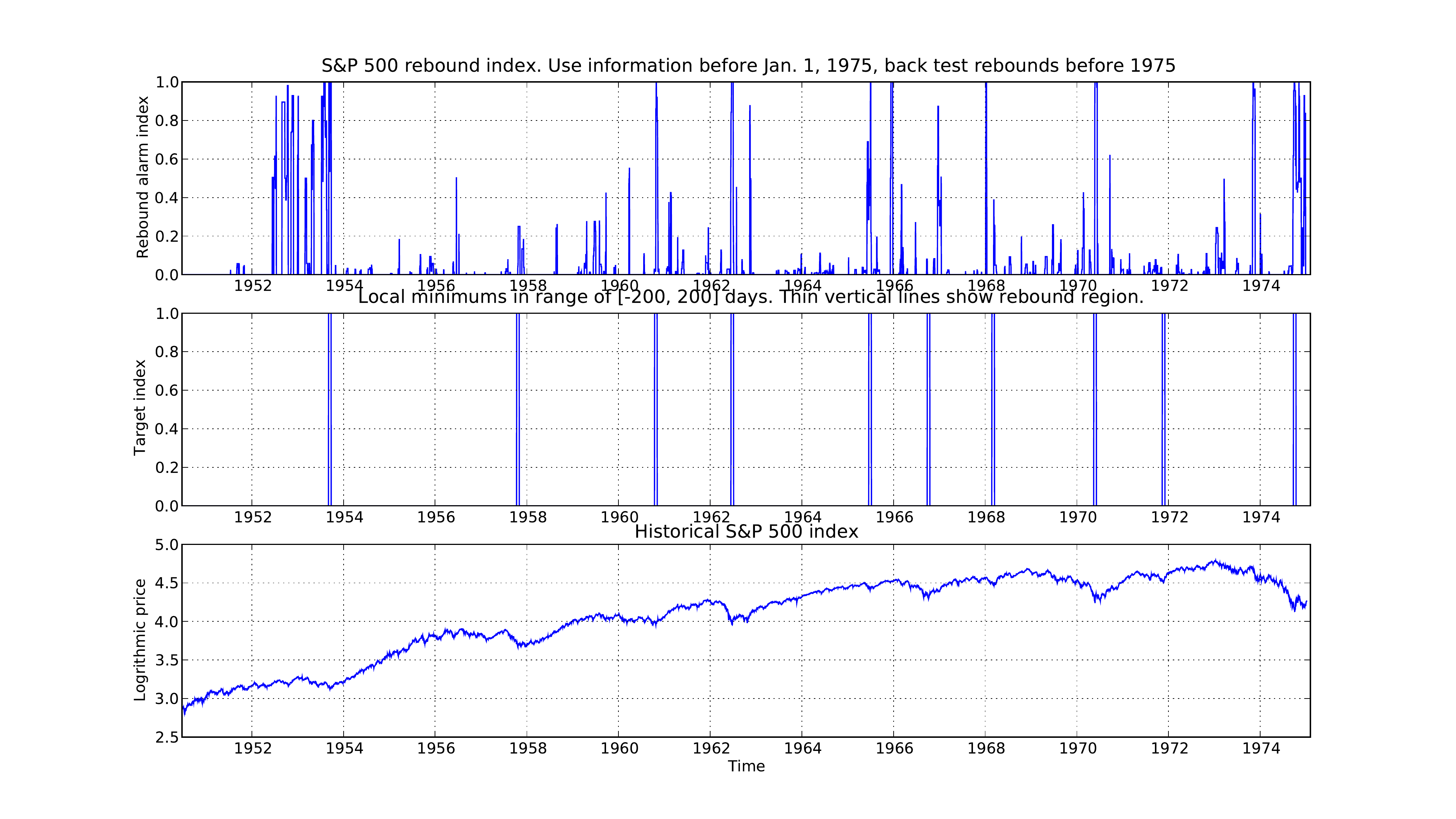}
\caption{Top panel: rebound alarm index of the back tests on each day of our dataset before 1975-01-01. The rebound alarm index is defined in [0,1]. The higher the rebound alarm index, the more likely the rebound will happen. Middle panel: the actual rebounds in this period, as defined by  (\ref{eq:rbd}). Lower panel: log-price of the S\&P 500 index.} \label{fig:bt}
\end{figure}

We use the features generated from the learning set, i.e. the fits before 1975-01-01, get the rebound alarm index before 1975-01-01 to check how the index performs. Fig.~\ref{fig:bt} shows the results. In this figure, the top panel shows the rebound alarm index for each trading day. The middle panel shows the times at which the real rebounds  
defined in (\ref{eq:rbd}) actually occurred. And the bottom panel is the log-price of the S\&P 500 index. 
Fig.~\ref{fig:bt} shows that the rebound alarm index diagnosed the real rebounds quite well. A quantitation
of what is meant by ``quite well'' will be given for the out-of-sample analysis performed below.

\begin{figure}[htp]
\centering
\includegraphics[width=\textwidth]{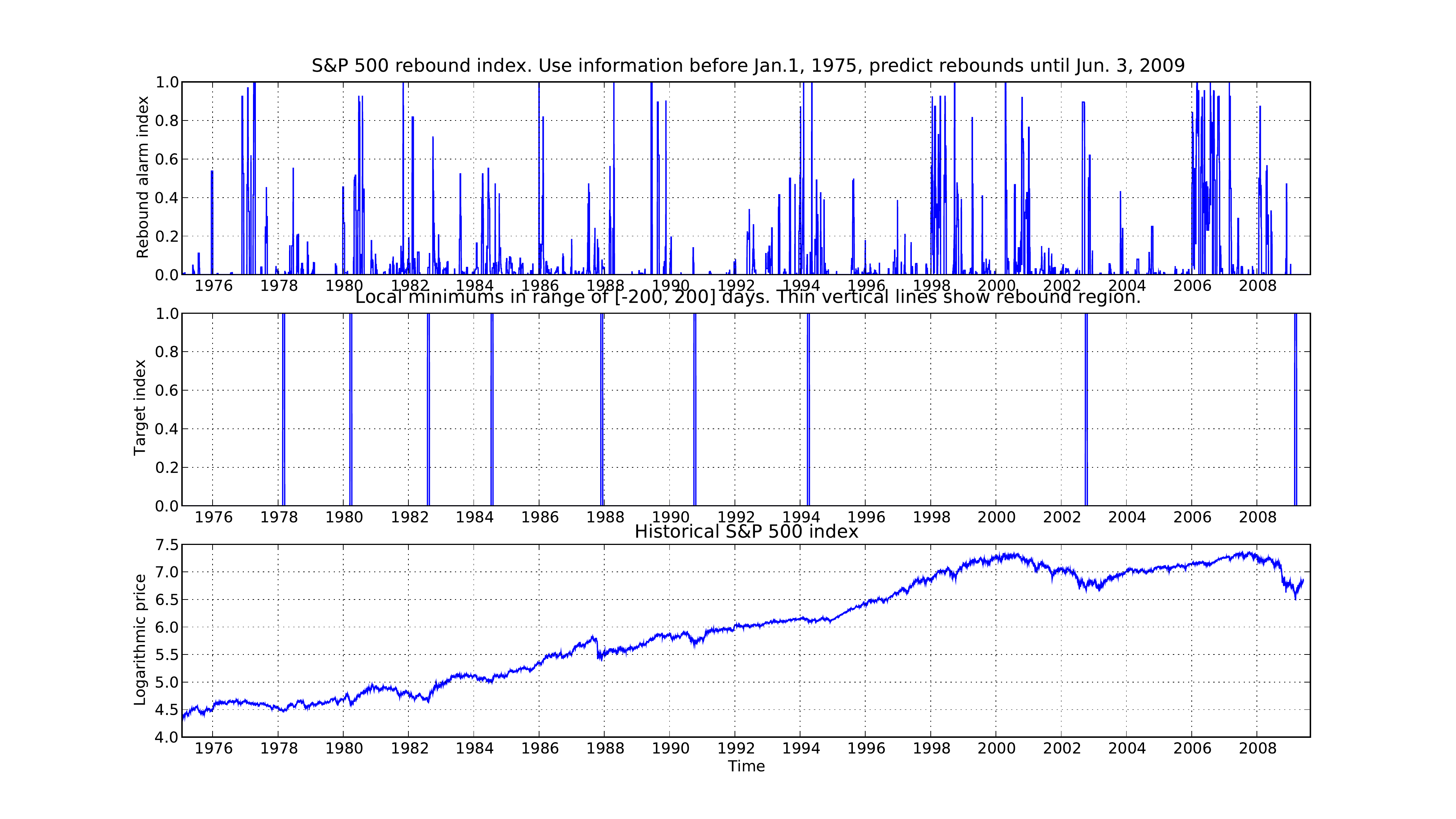}
\caption{Same as figure \protect\ref{fig:bt} for the out-of-sample prediction period after 1975-01-01.} \label{fig:predict}
\end{figure}

We now test how the rebound alarm index performs for ex-ante predictions. For this, we use the features learned from the learning set before 1975, then construct accordingly the rebound alarm index after 1975-01-01. The results are shown in Fig. \ref{fig:predict}. One can observe a satisfactory match between the periods when the rebound alarm index is large and the times when the rebounds actually occurred.

In order to quantify systematically and rigorously the quality of the performed predictions, we construct 
the associated error diagram as follows.
We set a threshold $T_h$ such that an alarm is declared when the rebound index
passes above  $T_h$.  When this occurs, the alarm is set to last for 40 additional days. 
The total alarm set is thus the set of all days for which the rebound index
is larger than $T_h$, augmented by the 40 consecutive days following the threshold days. 
A rebound is declared as having been predicted when it falls inside the total alarm set.
For a given threshold $T_h$, we determine the fraction of days which fall in the 
alarm set (alarm period / total period). We also count the fraction of rebounds which have not been predicted.
This defines the ``failure to predict'' equal to the fraction of missed rebounds. 
For a given threshold $T_h$, this gives one point in the plane where
``failure to predict'' is plotted as a function of the (alarm period / total period).
Changing $T_h$ from large to small values yields a line in the error diagram,
which quantifies the quality of the predictor in the way it addresses the trade-off
between never missing a rebound target and declaring the smallest number of alarms
(not ``crying wolf'' too much). 

It is clear that, if an alarm is declared at all times, no rebound will be missed.
This corresponds to the point $(1, 0)$. In contrast, declaring zero alarm
will miss all targets. This corresponds to the point $(0,1)$. In between, it 
is easy to see that the anti-diagonal line $y = 1 - x$ corresponds to 
random predictions, such that the fraction of successfully predicted events
is just the fraction of alarm time.
In contrast, an ideal perfect predictor corresponds to the convergence to the origin $(0,0)$,
for which there is no failure to predict any rebound, while using an asymptotically
vanishing fraction of time in alarms. A prediction system better than random
falls below the anti-diagonal line $y = 1 - x$.

\begin{figure}[htp]
\centering
\includegraphics[width=\textwidth]{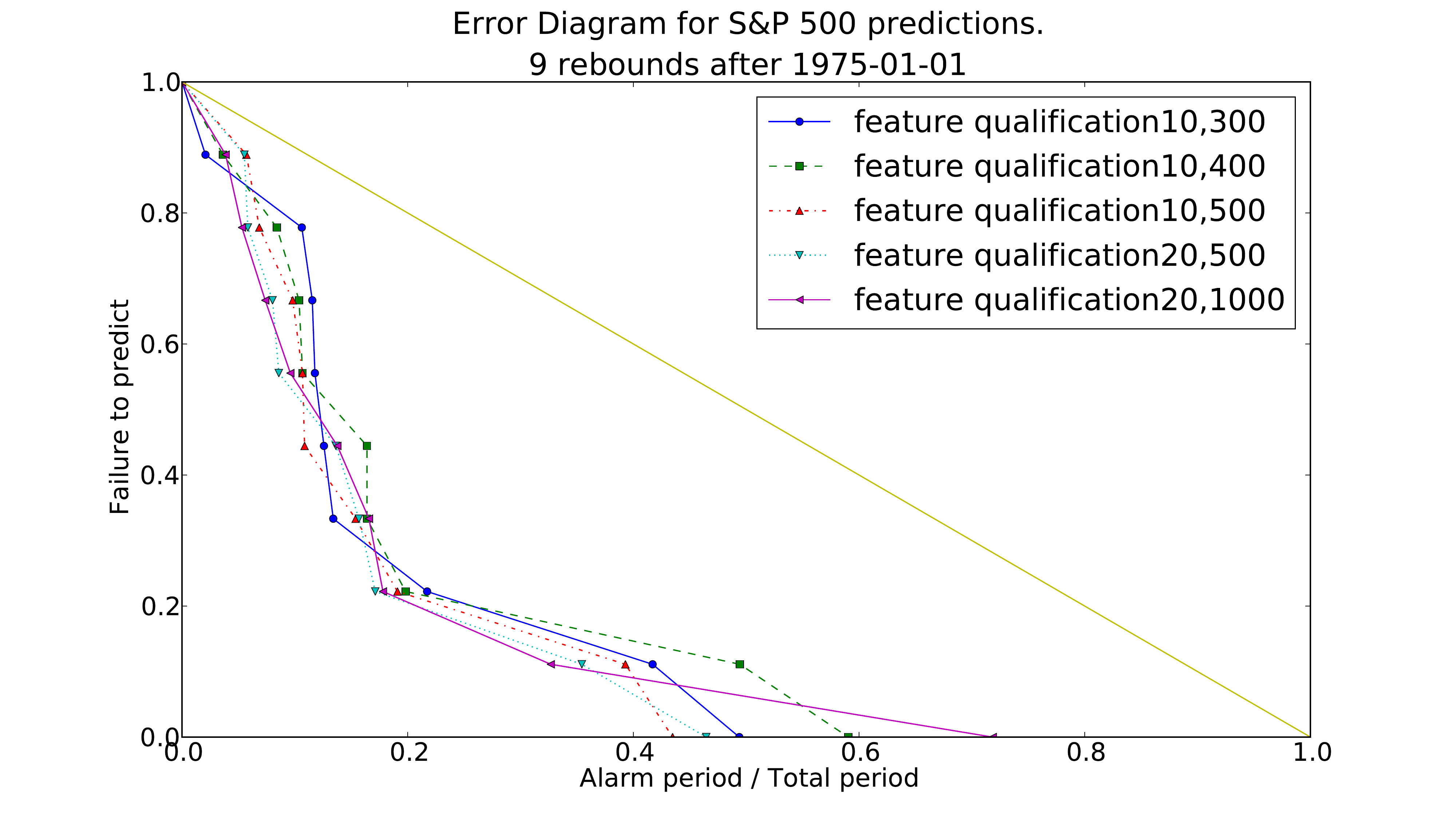}
\caption{Error diagrams for predictions performed after Jan. 1, 1975 with different types
  of feature qualifications. Feature qualification $\alpha, \beta$ means that when the
  occurrence of a certain trait in Class I is more than $\alpha$ and less
  than $\beta$ in Class II, then we call this trait a feature of Class I. }
\label{fg:22p}
\end{figure}

\begin{figure}[htp]
\centering
\includegraphics[width=\textwidth]{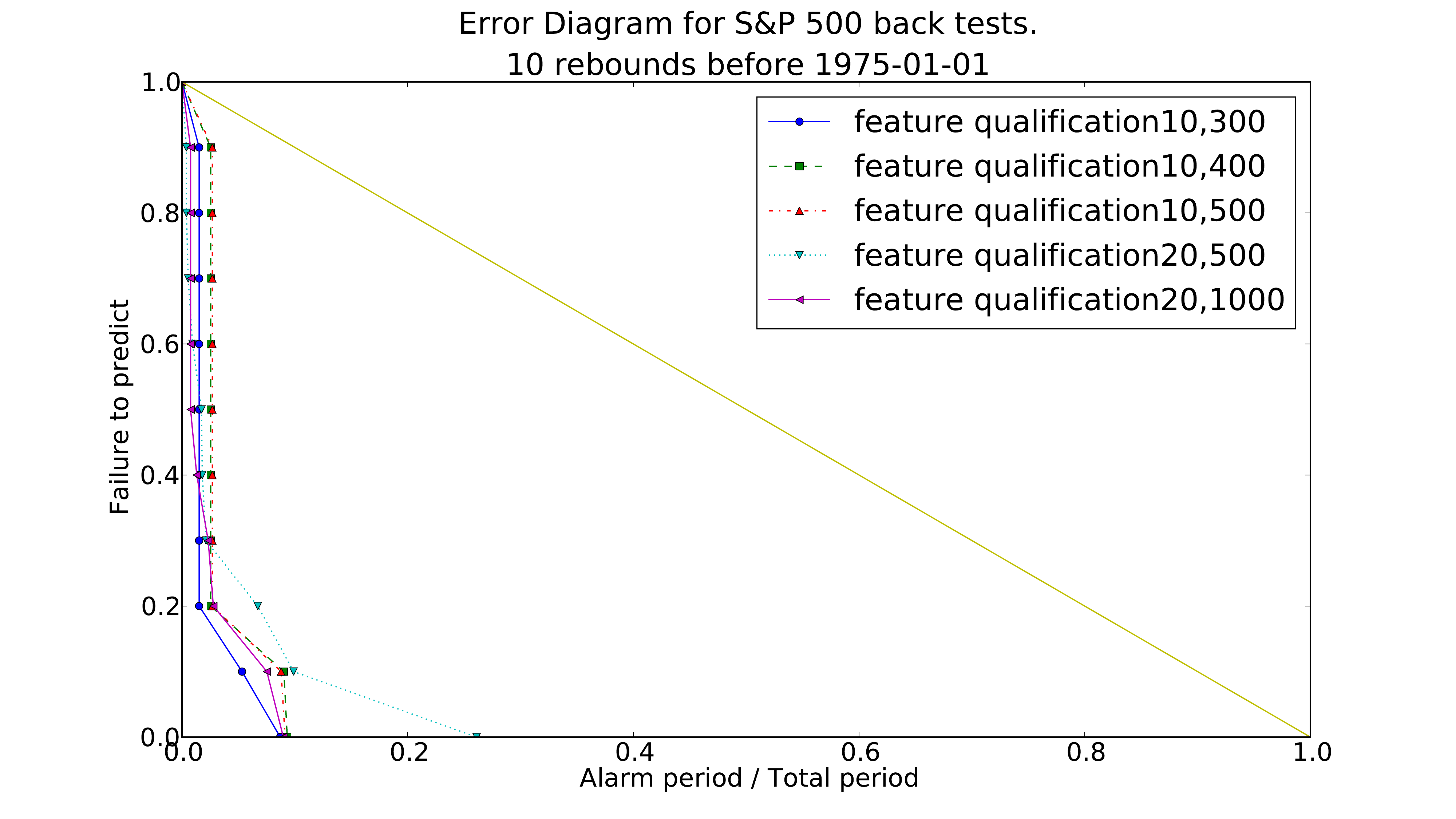}
\caption{Same as figure \protect\ref{fg:22p} for the back tests
performed before Jan. 1, 1975.}
\label{fg:22b}
\end{figure}

Fig. \ref{fg:22p} presents the error diagrams obtained for different types of features.
Fig. \ref{fg:22b} also plots the error diagrams obtained in the back tests.
The fast drop of failure to predict when the (alarm period / total period) is increased
from zero is the signature of a very significant predictive ability.

\section{Conclusion}
\label{conclusion}

We have presented partial results of an on-going ambitious
project, that aims to test scientifically, rigorously and systematically
the hypotheses that certain anomalous stock market regimes can
be diagnosed in real-time and their tipping point can be determined
better than chance. Specifically, we have presented two case
studies of the recent bubble on the Oil price that culminated
in early July 2008 and on the Shanghai stock market that 
crashes in August 2009. We have then shown how the proposed
log-periodic power law (LPPL) model can be extended
to so-called ``negative'' bubbles, these bearish market regimes
where the price spirals down in an accelerated way, before 
rebounding. We have presented statistical tests in- and out-of-sample
that demonstrate the strong predictive power of the proposed
methodology.

The research presented here is
highly unconventional in financial economics and will swim against the convention that bubbles cannot
be diagnosed in advance and crashes are somehow inherently impossible.  But as
Einstein once said: ``Problems cannot be solved at the same level of awareness
that created them.''  We thus propose a kind of Pascal's wager: is it really a
big risk for the community to 
explore the possibility of changing the conventional wisdom and open new
directions for the diagnostic of bubbles, ones that may eventually lead to
important policy and regulatory implications?

This research may have indeed global impacts. We confront directly the wide-spread belief
that crises are inherently unpredictable. If one can convince that some crises
can be diagnosed in advance and, what is even more important, if one can quantify the
associated uncertainties, this may help economists and policy makers
develop new approaches to deal with financial and economic crises.

But piling up more and more case studies, we are convinced that
the wealth of empirical results will foster many new theoretical insights
for the understanding of financial bubbles. We envision that this will have 
a natural spill-over to the questions of
how systemic risk and crisis depend on factors such as the
increasing interdependencies in the global economic systems 
that arise from the development of new financial instruments.
Central Banks, for example, could benefit from the models discussed here with a wide 
database of the national banks and of the firms and
calculate the risks associated with these various economic agents. 
At a microeconomic level, banks could implement such
models to supervise the activity of their branches to optimize risk management. 
The relatively small number of variables and
parameters involved as well as the relative simplicity of the mathematics involved ensures the practical applicability of the model in any framework of network dependence and interactions. 
Banks could find it useful to monitor the exposition of
correlation derivatives issued to immunize portfolios of liabilities. Financial institutions as well could appreciate the flexibility and real-time updating ability of our models to manage risk associated with large portfolios of assets.












\end{document}